\documentclass[final,5p,10pt,a4paper,times,sort&compress]{elsarticle}
\usepackage{ecrc}
\volume{371}
\firstpage{173}
\journalname{Physics Letters A}
\runauth{N. Karjanto, E. van Groesen}
\jnltitlelogo{\footnotesize PHYSICS LETTERS A}
\renewcommand{\figurename}{Fig.}
\usepackage{latexsym}
\usepackage{graphicx}
\usepackage[fleqn]{amsmath}
\usepackage{amsfonts}
\usepackage{amssymb}
\usepackage{epsfig}
\usepackage{subfigure}
\usepackage{microtype}
\usepackage{multicol}
\usepackage{balance}
\newtheorem{theorem}{Theorem}

\newtheorem{proposition}[theorem]{Proposition}

\newenvironment{proof}[1][Proof]{\textbf{#1.} }{\qquad {\scalebox{0.8}{$\square$}}} 
\usepackage{etoolbox}
\AtBeginDocument{%
  \patchcmd{\MaketitleBox}{\vspace*{-20pt}\fi}{\fi}{}{}%
}

\usepackage{fancyhdr}
\begin{document}
\begin{frontmatter}
\title{Note on wavefront dislocation in surface water waves}
\author{Natanael Karjanto\corref{cor1}}
\ead{n.karjanto@math.utwente.nl}
\cortext[cor1]{Corresponding author.}
\author{E. van Groesen}
\address{Department of Applied Mathematics, University of Twente, PO Box 217, 7500 AE, Enschede, The Netherlands \vspace{3pt}\\ 
{\upshape Received 3 January 2007; received in revised form 7 June 2007; accepted 7 June 2007} \vspace{3pt}\\
{\upshape Available online 29 June 2007} \vspace{3pt}\\
{\upshape Communicated by V.M. Agranovich}}

\begin{abstract}
At singular points of a wave field, where the amplitude vanishes, the phase may become singular and wavefront dislocation may occur.
In this Letter, we investigate for wave fields in one spatial dimension the appearance of these essentially linear phenomena. 
We introduce the Chu--Mei quotient as it is known to appear in the `nonlinear dispersion relation' for wave groups as a consequence of the nonlinear transformation of the complex amplitude to real phase-amplitude variables. 
We show that unboundedness of this quotient at a singular point, related to unboundedness of the local wavenumber and frequency, is a generic property and that it is necessary for the occurrence of phase singularity and wavefront dislocation, while these phenomena are generic too. 
We also show that the `soliton on finite background', an explicit solution of the NLS equation and a model for modulational instability leading to extreme waves, possesses wavefront dislocations and unboundedness of the Chu--Mei quotient. \\
{\small \copyright \ 2007 Elsevier B.V. All rights reserved.}
\end{abstract}

\begin{keyword}
Vanishing amplitude \sep Wavefront dislocation \sep Phase singularity \sep Chu--Mei quotient \sep Nonlinear dispersion relation \sep Soliton on finite background \vspace{5pt}\\

\PACS 46.40.Cd \sep 47.54.Bd \sep 47.35.Bb \sep 47.35 Fg \sep 94.05.Pt \sep 05.45.Yv \sep 52.35.Mw
\end{keyword}
\end{frontmatter}

\section{Introduction}

Many studies in the literature are dedicated to phenomena related to phase singularity and wavefront dislocation. 
Since generally both phenomena occur simultaneously, the term phase singularity is more often used in physical optics to describe what we will refer to as wavefront dislocation, for instance in~\cite{1Balistreri00}.
Alternatively, phase singularities also called intensity zeros, topological charges, or optical vortices~\cite{2Nye99,3Berry00,4Berry01,5Dennis04,6Soskin01}. 
In water waves, the disappearance of waves in a modulated train of surface gravity waves was described in~\cite{7Tanaka95}.

`Dislocation' is known for a long time in the field of material science. 
There it is used to describe an irregularity within a crystal structure, often responsible for the plastic deformation of metals and other crystalline solids. 
The concept was introduced as early as 1934 and proposed independently by~\cite{8Orowan34,9Polanyi34,10Taylor34}. 
Other references to dislocation in crystals are~\cite{11Hull01,12Dieter88,13Honeycombe84,14Read53}. 
The above phenomena are also found in several other branches of physics. 
A simple example of phase singularity is the singular time zone at the north pole~\cite{15Dennis01}. 
Experimental observations in a neon discharge in two-dimensional space-time are reported by Kr\'{a}sa~\cite{16Krasa81}. 
A study of it in the Aharonov--Bohm effect is done by Berry~\cite{17Berry80}. 
An analysis for constructing a theory of wavefront dislocation using catastrophe theory is developed by Wright~\cite{18Wright79}. 
A study of the phenomenon in optics, particularly in monochromatic light waves is reported in~\cite{19Basistiy95}.

Extensive references about many topics related to dislocations from theoretical to experimental observations and applications can be found in~\cite{20Nabarro04}. 
Line singularities in vector and electromagnetic waves, including the paraxial case, when waves propagate in a certain direction and the general case,\, when waves propagate in all directions, is discussed in~\cite{21Nye97}. 
A theoretical framework for understanding the local phase structure and the motion of the most general type of dislocation in a scalar wave, how this dislocation may be categorized and how its structure in space and time is related, has been studied by Nye~\cite{22Nye81}. 
Statistical calculations associated with dislocations for isotropically random Gaussian ensembles, that is, superpositions of plane waves equidistributed in direction but with random phases, are given in~\cite{3Berry00}. 
Knotted and linked phase singularities in monochromatic waves by constructing exact solutions of the Helmholtz equation are given in~\cite{4Berry01}.

Apparently, the same phenomenon is also observed in 3D surfaces of constant phase (wavefronts) of a wave field. 
A new concept of `wavefront dislocation' was introduced in 1974 by Nye and Berry~\cite{23Nye&Berry74} and is used to explain the experimentally observed the appearance and disappearance of crest or trough pairs in a wave field. 
In their paper, the examples given are in two and three space dimensions plus time. Other terminologies that are also often used to describe the phenomenon are death and birth of waves, and annihilation and creation of waves. 
When dealing with waves, Nye and Berry~\cite{23Nye&Berry74} showed that dispersion is not really involved when wavefront dislocation occurs, while, on the other hand, Trulsen~\cite{24Trulsen98} explained that wavefront dislocation is a consequence of linear dispersion alone and predicted by the linear Schr\"{o}dinger equation, an example of paramount importance of a linear dispersive wave equation.

In this Letter, we restrict to wave fields with one spatial and one temporal variable. 
Even for this simplest case, we sensed some confusion in the cited references above about the equivalence of the possibly different phenomena of phase singularity and wavefront dislocation. Moreover, it was not very clear if these phenomena are exceptional, rare events or should be expected at any point of vanishing amplitude. 
From a more practical point of view, we wanted to use the appearance of wavefront dislocations that we had found in the theoretical expression of a `soliton on finite background' as a check-in measured signals of waves that were generated in a hydrodynamic laboratory. 
Robustness of such a phenomenon for perturbations of various kind is then required, a result that was not found in the cited references.

This Letter is organized as follows. 
In Section~\ref{Preliminaries}, we present the basic notions of phase singularity, wavefront dislocation and the Chu--Mei quotient that will be used in this paper. 
Further, we give the most trivial examples of surface wave fields, namely superpositions of two and three monochromatic waves. 
With these examples, we will show that already for trichromatic waves, phase singularity and wavefront dislocation will be generic properties, but also that phase singularity is not necessarily accompanied by wavefront dislocation. 
In Section~\ref{WDwavegroups} we study these aspects for wave groups. 
We will show that unboundedness of the Chu--Mei quotient is a necessary condition for the occurrence of wavefront dislocation. 
A perturbation analysis is done to show that boundedness of the Chu--Mei quotient is an exceptional case.
Although all these phenomena are essentially linear, in Subsection~\ref{SubsectionSFB} we investigate these phenomena for the special solution of NLS, the soliton on finite background, SFB. The final section concludes the Letter with some conclusions and remarks.

\section{Preliminaries} \label{Preliminaries}

This section is devoted to collect preliminary definitions that will be used in this Letter to study wavefront dislocation. 
We also illustrate degenerate and generic cases of the phenomena by using simple wave fields that consist of a superposition of a few harmonic modes.
\vfill

\subsection{Basic notions} \vspace{5pt} 
\vfill

Let $\eta(x,t)$  be a real-valued function that describes a surface wave field in one space variable $x$ and time $t$. 
The complexification of $\eta$ is defined by the Hilbert transform $\mathcal{H}$$[\eta]$, given by $\eta_{\textmd{c}}(x,t) = \eta(x,t) + i \mathcal{H}[\eta(x,t)]$. 
Written in polar form with real-valued phase and amplitude variables we get $\eta_{\textmd{c}}(x,t) = a(x,t) e^{i\Phi(x,t)}$. 
The {\itshape local wavenumber} and {\itshape local frequency} are defined respectively as $k(x,t) = \partial_{x}\Phi$ and $\omega(x,t) = -\partial_{t} \Phi$.
\vfill

The phase $\Phi$ is uniquely defined for smooth functions $\eta$ for all $(x,t) \in \mathbb{R}^{2}$ for which the amplitude does not vanish. 
When the wave field has {\itshape vanishing amplitude}, i.e. if $a(\hat{x},\hat{t}) = 0$, we call $(\hat{x},\hat{t}) \in \mathbb{R}^{2}$ a {\itshape singular point}. 
In the Argand diagram (the complex plane), the time signal at a fixed position corresponds to an evolution curve $t \mapsto \eta_{\textmd{c}}(x,t)$; a singular point $(\hat{x},\hat{t})$ corresponds to an evolution curve that is at the origin of the complex plane.
\vfill

For nonzero amplitude, the phase of $\eta_{\textmd{c}}$ has a well-defined value, but at a singular point the phase $\Phi$ may be undetermined or even be singular. 
We will say that the wave field $\eta(x,t)$ has a {\itshape phase singularity} at the singular point $(\hat{x},\hat{t})$ if $\Phi(x,t)$ is not continuous. 
As is clear from the interpretation in the Argand diagram, in most cases the trajectory will cross the origin and the phase will be discontinuous and have a $\pi$-jump. 
Only in case the origin acts as a reflection point, the phase will be continuous. 
Examples are easily constructed for both cases by superposition of just a few waves.
\vfill

Wavefront dislocation is observed when waves at a certain point and time merge or split. 
Necessarily this can happen only at a singular point, as we will see. 
Formally we will define that the wave field $\eta(x,t)$ has {\itshape wavefront dislocation} of strength $n \neq 0$ in the area of the  $xt$-plane that is enclosed by a contour if the following contour integral has the given integer multiple of $2\pi$~\cite{23Nye&Berry74,25Berry81,26Berry98}:
\begin{equation}
\oint \mathrm{d}\Phi = \oint (k \, \mathrm{d}x - \omega \, \mathrm{d}t) = \iint \left(\frac{\partial \omega}{\partial x} + \frac{\partial k}{\partial t} \right) \, \mathrm{d}x \, \mathrm{d}t =  2n\pi,  \; n \neq 0.     \label{contour_integral1}
\end{equation}

Instead of taking an arbitrary closed curve, it is also possible to investigate the property of a given singular point. 
Then, by taking a circle of radius $\epsilon$, and allowing the radius to shrink to zero, the strength of the singular point is found from
\begin{equation}
I = \lim_{\epsilon \rightarrow 0} \oint_{C(\epsilon)} \mathrm{d}\Phi
  = \lim_{\epsilon \rightarrow 0} \int_{0}^{2\pi} \frac{\mathrm{d}\Phi}{\mathrm{d}\theta} \, \mathrm{d}\theta, 	    \label{contour_integral2}
\end{equation}
where $\theta$ is the angle variable describing the circle. 
If $I = 0$ there is no wavefront dislocation, while if $I = \pm 2\pi$ there is wavefront dislocation. 
More specifically, splitting of waves for progressing time will occur if for increasing $x$ the value of $I = -2\pi$; for $I = 2\pi$, merging of waves will happen for increasing~$x$.

\newpage
For a full description of the appearance of wavefront dislocation it turns out to be useful to introduce another quantity. 
This is the so-called `Chu--Mei quotient'\footnote[1]{Some authors call this quotient the `Fornberg--Whitham term'~\cite{27Infeld90}, referring to~\cite{28Fornberg78}. 
However, throughout this Letter we call it the `Chu--Mei quotient', since they introduced it for the first time~\cite{29Chu70,30Chu71} when they derived the modulation equations of Whitham's theory~\cite{31Whitham67} for slowly varying Stokes waves. 
However, the quotient already appeared earlier in~\cite{32Karpman67,33Karpman69} when they consider modulated waves in nonlinear media.} defined for signalling problems by
\begin{equation}
  \textmd{CMq} = \frac{\partial^{2}_{t}a}{a};
\end{equation}
for initial value problems, the derivative with respect to $x$ is used instead of to $t$. 
This quotient appears in the dispersion relation for both linear and nonlinear dispersive wave equations and has a clear interpretation in this context, as we will show in Section~\ref{WDwavegroups}. 
\begin{figure}[htbp]
\twocolumn[{%
\begin{center}%
\subfigure[]{\includegraphics[width=0.31\textwidth]{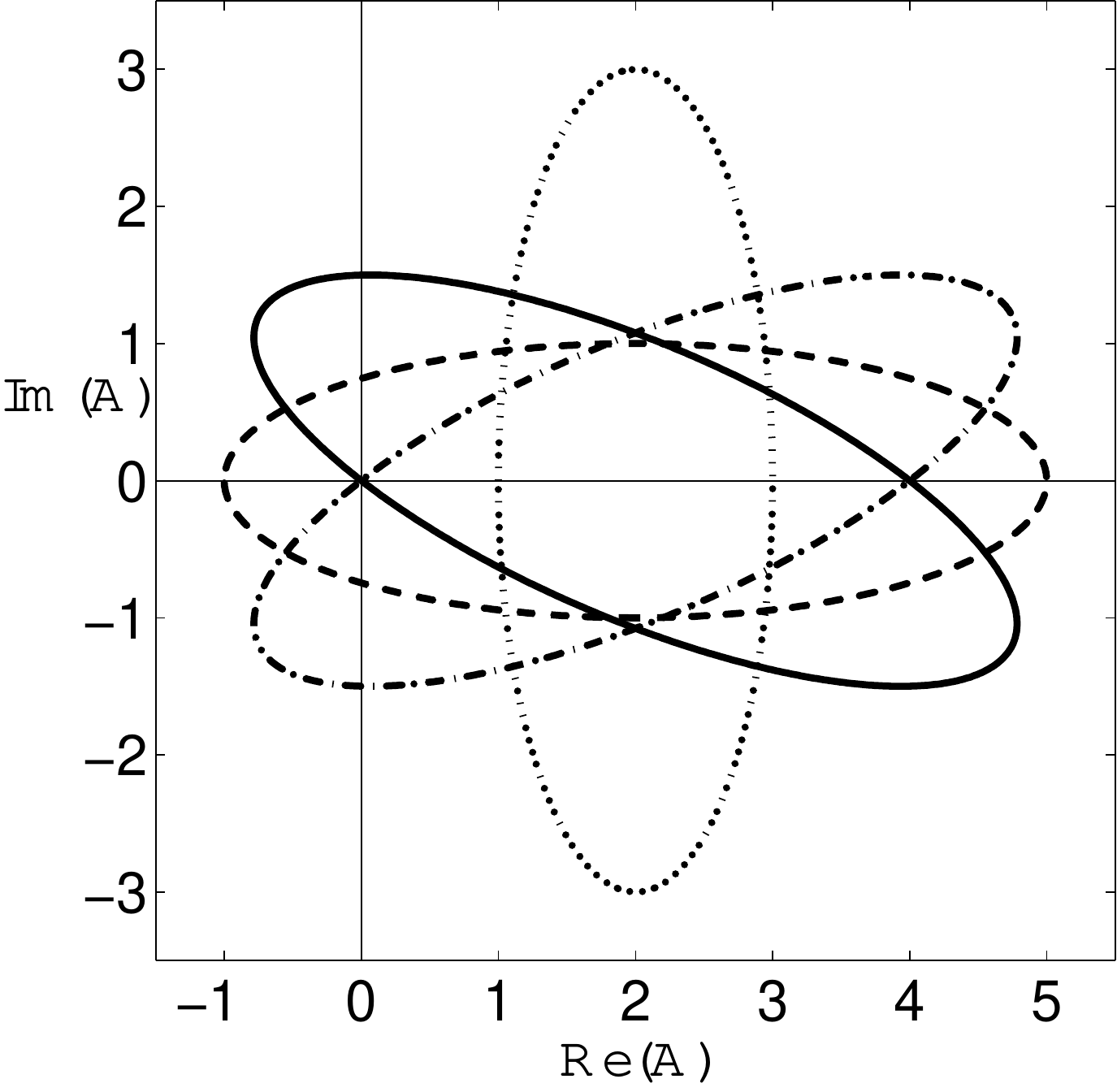}} \hspace{2cm}
\subfigure[]{\includegraphics[width=0.4\textwidth]{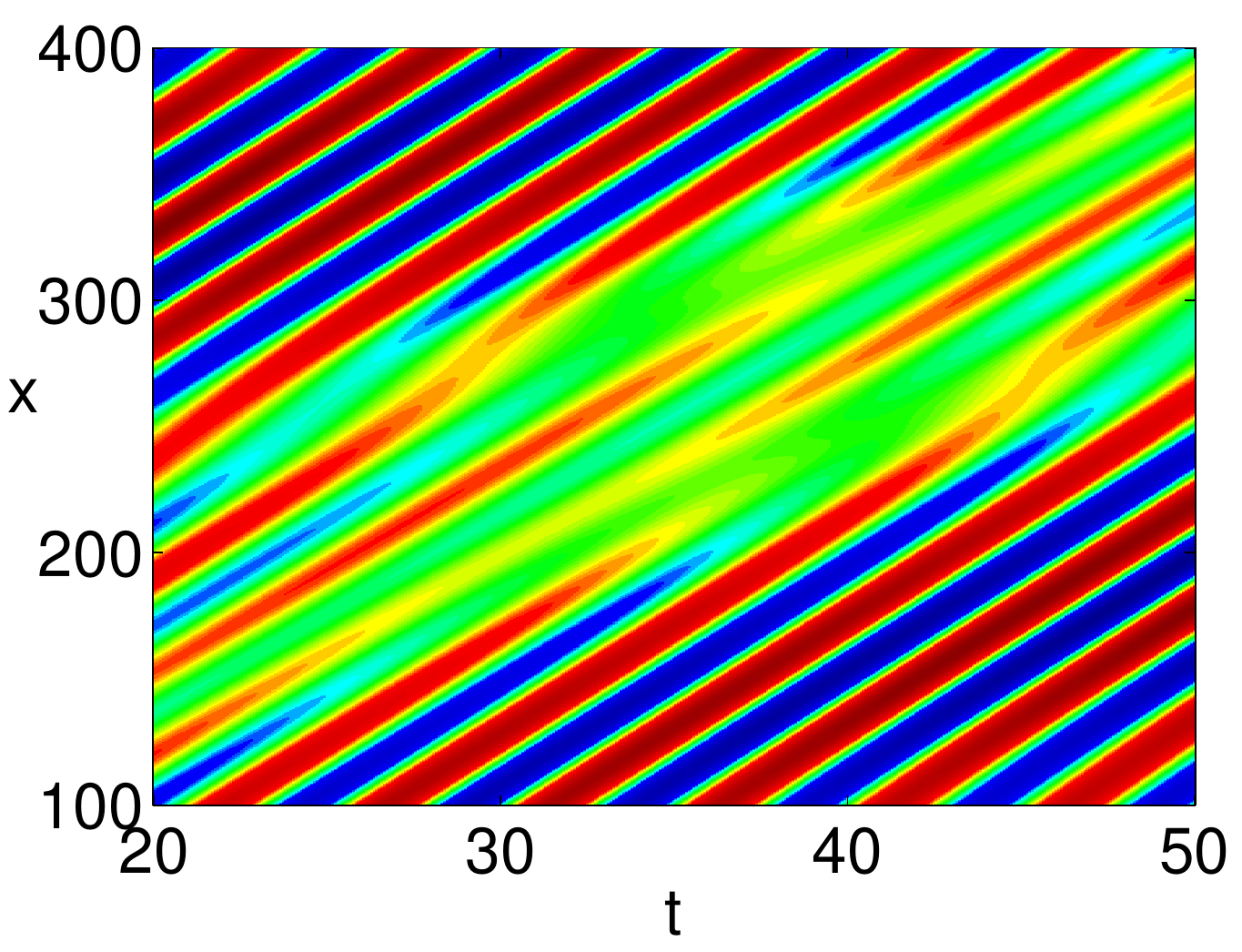}}
\end{center}
\caption{\footnotesize At the left, the evolution in the Argand diagram is shown for the trichromatic wave parameterized by $\tau$ and plots are given for different values of $\xi$: 
$\xi = 0$ (dotted); $\xi = \hat{\xi}_{1}$ (solid); $\xi = \pi/(2\nu^{2})$ (dashed); $\xi = \hat{\xi}_{2}$ (dash-dot). 
The evolution curves are counterclockwise ellipses and follow a clockwise direction for increasing~$\tau$. 
When the ellipse crosses the origin, a phase singularity and wavefront dislocation occurs. 
At the right, the density plot is shown of the trichromatic wave near a phase singularity where splitting and merging of waves can be seen.} \label{trichromatic} \vspace{10pt}
}]  
\end{figure}

\vspace*{-20pt}
In the remainder of this Letter, we will show the following inclusive relations between the concepts introduced above. 
In Section~\ref{WDwavegroups} we will show that unboundedness of the Chu--Mei quotient is a necessary condition for wavefront dislocation to occur in wave groups. 
Further, wavefront dislocation at a point implies that there is phase singularity and phase singularity can only occur at singular points. 
Although we give examples that the reversed implications are not valid, this will only happen for degenerate cases. 
\emph{Generically} it will be the case that at a singular point there is phase singularity, wavefront dislocation and unboundedness of the Chu--Mei quotient.
\vspace*{10pt}

A monochromatic wave will not have any singular point,~and therefore it is not interesting for our investigations. 
A bichromatic wave can have singular points, which may have phase singularity. 
In that last case, there will be no wavefront dislocation and the Chu--Mei quotient will be finite. 
A combination of three monochromatic waves can show all the phenomena; we will briefly describe these illustrative cases in the following subsection.

\subsection{Bichromatic and trichromatic wave fields} 	\label{Bichromatic}

Consider the superposition of two monochromatic waves, {\spaceskip 0.6 em \relax known as the bichromatic waves.
A complexified form is} $\eta_{\textmd{c}}(x,t) = |A_+| \exp(i\theta_+) + |A_-| \exp(i\theta_-)$, where $A_\pm = |A_\pm|e^{i\phi_\pm}$ are the complex-valued amplitudes and where 
$\theta_\pm = k_\pm x - \omega_\pm t + \phi_\pm$ are the phases of the constituent monochromatic waves. 
Assume that the waves satisfy a truly dispersive equation so that their phase velocities are different.
Inspection of the real amplitude of the superposition shows that singular points can only, and will, happen for $|A_+| = |A_-|$ and then for $\theta_+ -\theta_- = \frac{1}{2}n \pi$, $n \in \mathbb{Z}$.

{\spaceskip 0.6 em \relax The Chu--Mei quotient is bounded at a singular point:} $\textmd{CMq} = \lim_{t \rightarrow \hat{t}} \frac{\partial_{t}^{2}a}{a}(x = \hat{x},t) = - \nu^{2}$. 
Using Proposition~\ref{ChuMei2WD} below, this implies that this wave field does not have wavefront dislocation. 
Calculating the contour integral~\eqref{contour_integral2} around any singular point $(\hat{x},\hat{t})$ it is found that indeed $I(\hat{x},\hat{t}) = 0$.

Consider the superposition of three monochromatic waves. 
Already in the case, generically phase singularity and wavefront dislocation will occur whenever the amplitude vanishes. 
An example is a solution of the linear version of the NLS equation~\eqref{NLSE}. 
Namely, $\eta(x,t) = A_{\textsc{tc}}(\xi,\tau) e^{i(k_{0}x - \omega_{0}t)} + $ c.c., where $\xi = x$, $\tau = t - x/V_{0}$, $V_{0}$ is the group velocity, $A_{\textsc{tc}}(\xi,\tau) = \sum_{n = 0}^{2} b_{n} e^{i(\kappa_{n}\xi - \nu_{n}\tau)}$, where $b_{n} \neq 0$, $\kappa_{0} = 0 = \nu_{0}$, $\nu_{1} = \nu = -\nu_{2}$ and $\kappa_{1} = \nu^{2} = \kappa_{2}$.

\begin{figure*}[htbp]
\begin{center}
\includegraphics[width = 0.36\textwidth]{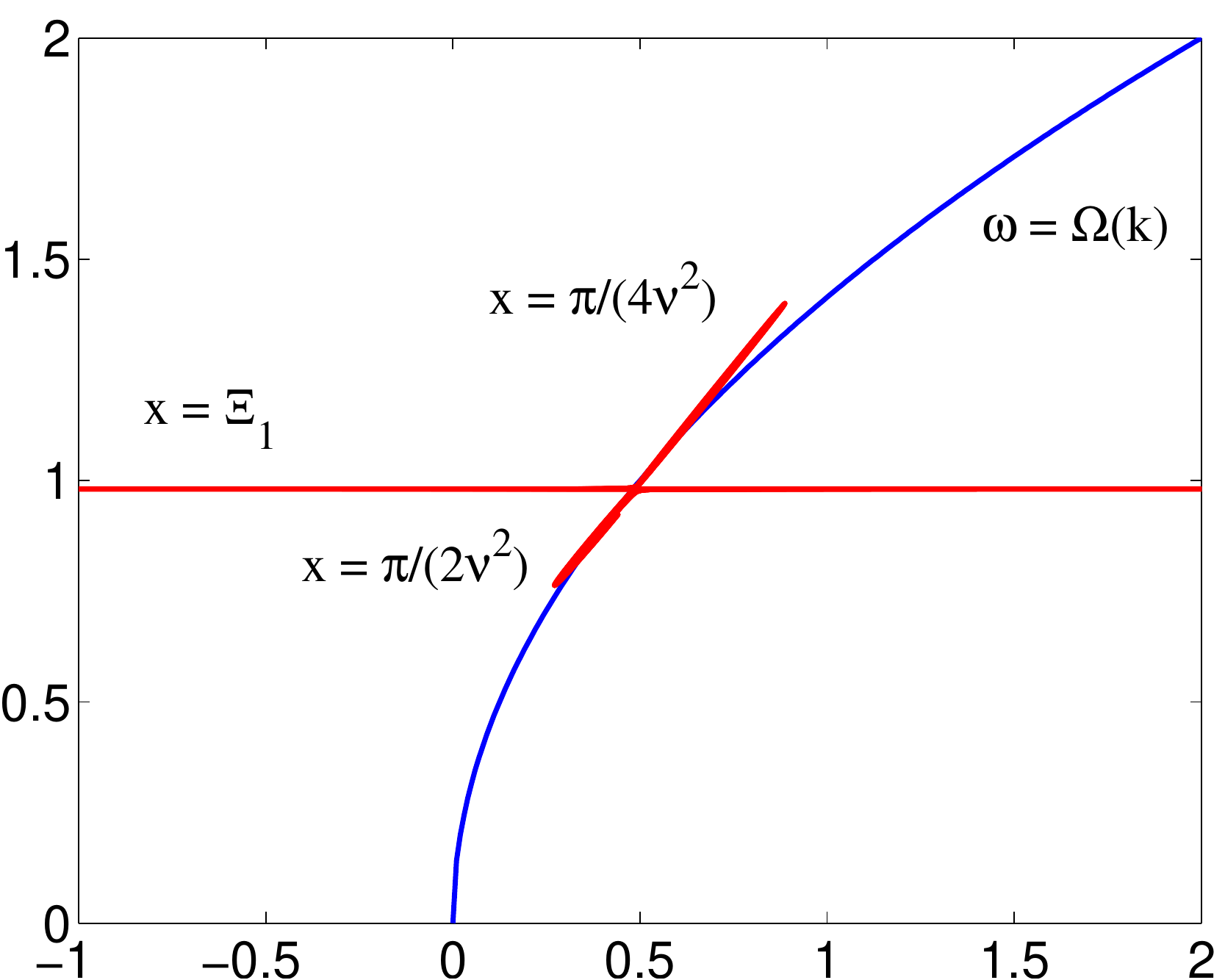}	\hspace{2cm}
\includegraphics[width = 0.36\textwidth]{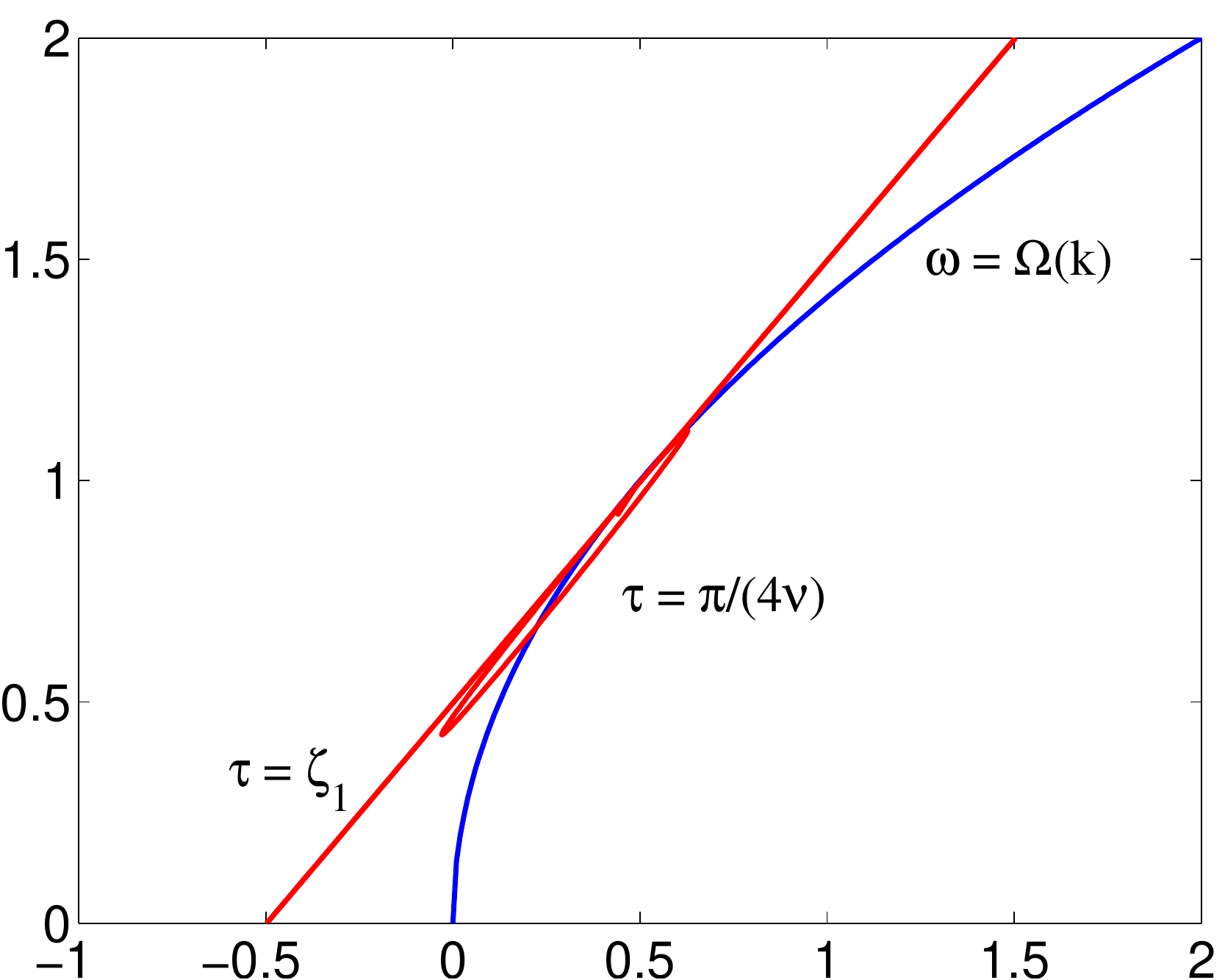}
\end{center}
\caption{\footnotesize Plots of the local wavenumber $k$ (horizontal axis) and the local frequency $\omega$ (vertical axis) in the dispersion plane of the considered trichromatic wave. In the left plot, for fixed position some trajectories are shown parameterized by the time, showing that the local wavenumber becomes unbounded at the instance of singularity; similarly, the right plot is for a fixed time with trajectories parameterized by position, showing the local frequency becoming unbounded at the singular position.}     \label{Linear_LWLF}
\end{figure*}

We illustrate some aspects for the case studied by Trulsen \cite{24Trulsen98}, for which $b_{0} = 2$, $b_{1} = -2$, and $b_{2} = 1$, $\nu = 1/13$ and $\omega_{0} = 1$. 
The motion of the amplitude in the complex plane, shown in Figure~\ref{trichromatic}(a), makes it clear that there are singular points with phase singularity. 
The appearance of wavefront dislocation is shown in the density plot in Figure~\ref{trichromatic}(b), and can be investigated in detail by counting the number of waves in one period. 
The Chu--Mei quotient is unbounded at the singular points.
This is related to the fact that the local frequency and local wavenumber become unbounded, as shown in Figure~\ref{Linear_LWLF}.

\section{Wavefront dislocations in wave groups}			\label{WDwavegroups}

In the previous section, we showed that already a superposition of three monochromatic waves can show wavefront dislocation at singular points. 
In this section, we will consider linear and nonlinear dispersive wave equations and show that a necessary condition for a wave group to have a wavefront dislocation is that the Chu--Mei quotient is unbounded. 
Moreover, we will also show that the unboundedness of this term is a generic property: if it is bounded at a singular point for an exceptional case, any perturbation of the waves will result into an unbounded Chu--Mei quotient.

\subsection{Linear and nonlinear dispersive wave equations}

We consider a linear or nonlinear dispersive wave equation. As a model for mainly unidirectional propagation, we can take an evolution equation of KdV type:
\begin{equation}
\partial_{t}\eta + i \Omega(-i\partial_{x})\eta + \partial_{x} N(\eta) = 0. \label{LDWE}
\end{equation}
Here $k \mapsto \Omega(k)$ determines the dispersion relation; the inverse will be denoted by $K$: $K = \Omega^{-1}$. 
The weak nonlinearity is given by $N(\eta) = a \eta^{2} + b \eta^{3}$, but is of little relevance for the following discussion about wavefront dislocation as we shall see, so taking a linear equation for which $a = 0 = b$ is possible.

When looking for a wave group with carrier frequency $\omega_{0}$, the evolution is described with a complex amplitude $A$, and is then given in lowest order by
\begin{equation*}
\eta(x,t) = \epsilon \, A(\xi,\tau)e^{i\theta_{0}} + \textmd{c.c.},
\end{equation*}
where $\theta_{0} = k_{0}x - \omega_{0}t$, with $k_{0} = K(\omega_{0})$ \ and c.c. denotes~the complex conjugate of the preceding term. 
The amplitude is described in a time-delayed coordinate system: $\xi = x$ and $\tau = t - x/V_{0}$ where $V_{0} = \Omega'(k_{0}) = 1/K'(\omega_{0})$. 
This transformation is suitable for studying the evolution in space, for the signalling problem. 
The resulting equation for $A$ is then the spatial nonlinear Schr\"{o}dinger (NLS) equation, given by
\begin{equation}
\partial_{\xi}A + i \beta \partial_{\tau}^{2} A + i\gamma |A|^{2}A = 0. \label{NLSE}
\end{equation}
Here $\beta = - \Omega''(k_{0})/(2[\Omega'(k_{0})]^{3})$ is related to the group velocity dispersion, while $\gamma$ is a transfer coefficient from the nonlinearity ($\gamma = 0$ for the linear equation).

By writing $A$ in its polar form with the real-valued amplitude $a$ and the real-valued phase $\phi$, $A = a(x,t)e^{i\phi(x,t)}$, and substituting into~\eqref{NLSE}, we obtain the coupled phase-amplitude equations. 
In the original physical variables, the `energy equation' is given by
\begin{equation*}
\partial_{x}(a^{2}) + \partial_{t}[K'(\omega) a^{2}] = 0,
\end{equation*}
and the phase equation can be written as the nonlinear dispersion relation:
\begin{equation}
K(\omega) - k = \beta \frac{\partial_{t}^{2}a}{a} + \gamma a^{2}. \label{linearChuMei}
\end{equation}
Even for a linear equation, the phase equation~\eqref{linearChuMei} contains an additional nonlinear term which results from the fact that the transformation $A \mapsto (a,\phi)$, $A = a e^{i\phi}$ itself is nonlinear.

At vanishing amplitude, the term from nonlinearity of the equation vanishes, $\gamma a^{2} = 0$, which shows that the nonlinearity does not play an important role at vanishing amplitude and hence for the phenomena to follow. 
Only the Chu--Mei quotient plays a significant role in understanding phase singularity and wavefront dislocation phenomena. 
Unboundedness of the Chu--Mei quotient implies that $K(\omega) - k$ becomes unbounded, and hence that the local wavenumber and the local frequency become unbounded.

Now we will show that the wavefront dislocation can only appear if the Chu--Mei quotient is unbounded.
\begin{proposition}
A necessary condition for a wave field to have a wavefront dislocation at a singular point is that the Chu--Mei quotient is unbounded.		  \label{ChuMei2WD}
\end{proposition}
\begin{figure*}[htbp]
\begin{center}
\includegraphics[width = 0.36\textwidth]{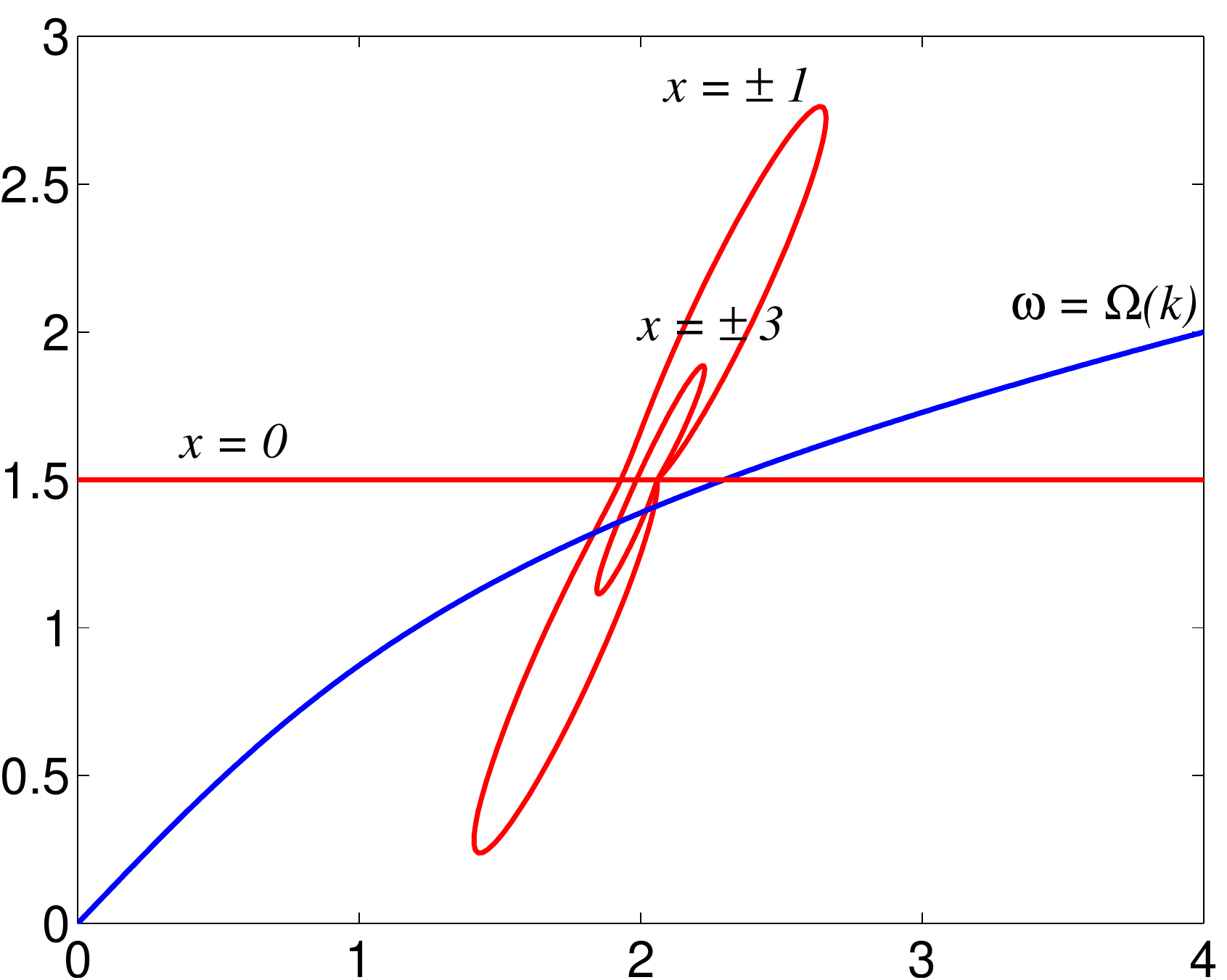}	\hspace{2cm}
\includegraphics[width = 0.36\textwidth]{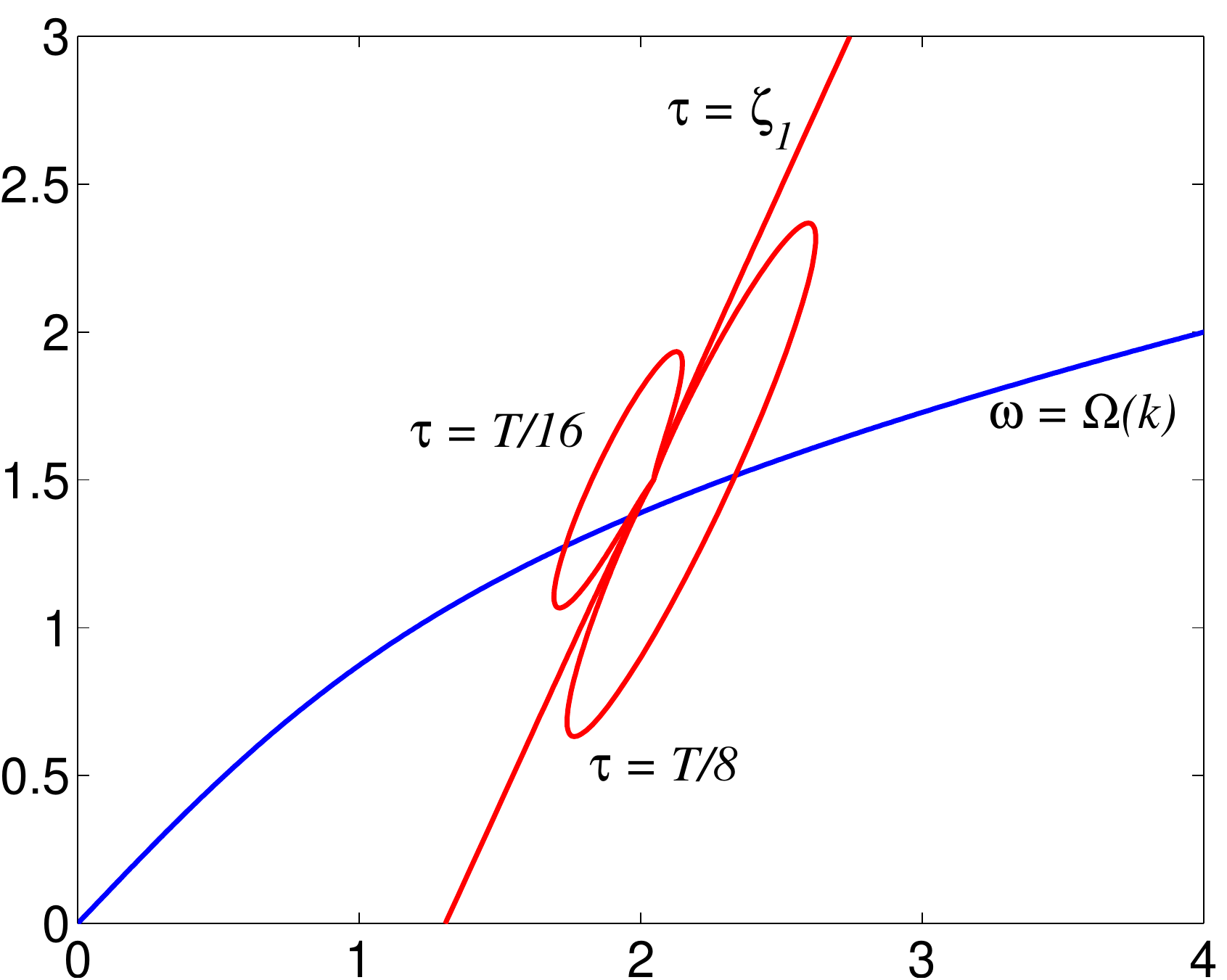}
\caption{\footnotesize Plots of the local wavenumber $k$ (horizontal axis) and the local frequency $\omega$ (vertical axis) in the dispersion plane for $\nu = \frac{1}{2}$.
At $x = 0$, the local wavenumber becomes unbounded (left), and at $\tau = \zeta_{1}$, the local frequency becomes unbounded (right).}	    \label{SFB_LWLF}
\end{center}
\end{figure*}
\begin{proof}
The proposition means that if the contour integral $d\phi$ is nonzero, then the Chu--Mei quotient is unbounded at the singular points. 
We will show its contraposition, namely if the Chu--Mei quotient is bounded at singular points, then the contour integral vanishes and there is no wavefront dislocation.
The Chu--Mei quotient is bounded at a singular point means that either both local wavenumber and local frequency are bounded at the singular point or that the local wavenumber or local frequency is unbounded, but $|K(\omega(x,t)) - k(x,t)| < \infty$.
For the first case, since both quantities are bounded, the integrand in the contour integral~\eqref{contour_integral1} is bounded, and hence vanishes in the limit for vanishing contour around the point.
For the latter case, it means that there exists a positive constant $M$ such that $K(\omega(x,t)) - M \leq k(x,t) \leq K(\omega(x,t)) + M$.
Hence, since $K(\omega) \rightarrow \pm \infty$ if and only if $\omega(x,t) \rightarrow \pm \infty$, both wavenumber and frequency have to be unbounded. 
For evaluating the contour integral~\eqref{contour_integral1}, observe that
\begin{align*}
& \oint(K(\omega) \, \mathrm{d}x - \omega \, \mathrm{d}t) - \oint M \, \mathrm{d}x \\
& \quad \leq \oint (k \, \mathrm{d}x - \omega \, \mathrm{d}t) \leq \oint(K(\omega) \, \mathrm{d}x - \omega \, \mathrm{d}t) + \oint M \, \mathrm{d}x.
\end{align*}
The contribution $\oint M \, \mathrm{d}x$ vanishes in the limit for shrinking contour, and the same holds for the integral $\oint(K(\omega) \, \mathrm{d}x - \omega \, \mathrm{d}t)$ by selecting a limiting contour such as a rectangle for which the length of the sides are chosen appropriately, for instance $\mathrm{d}x = \mathcal{O}(\omega/K(\omega)) \, \mathrm{d}t$.
Thus, also in this case the contour integral~\eqref{contour_integral1} vanishes, and there is no wavefront dislocation.
\end{proof}

\subsection{The Chu--Mei quotient under perturbation} 		\label{ChuMei}
\vspace{5pt}

We will now show that the boundedness of the Chu--Mei quotient at a singular point is exceptional: almost any perturbation of the wave field will make the quotient to become unbounded. 
This is intuitively clear by looking at the trajectory in the Argand diagram: at a singular point, the trajectory crosses the origin, $a = 0$, and it will be exceptional if it does this with vanishing `acceleration' $\partial_{t}^{2}a = 0$.
\vspace{5pt}

The translation of this result to complex-valued functions will give the required statement. 
Indeed, let $F: \mathbb{R}^{2}\rightarrow \mathbb{C}$, and denote by $F^{\prime}$ and $F^{\prime\prime}$ respectively the first and second derivative with respect to the parameter $t$ or, actually, in any direction. 
Then defining the amplitude $a$ as $a^{2} = |F|^{2}$, after some manipulations we get
\begin{equation*}
\frac{\partial_{t}^{2}a}{a} = \frac{\textmd{Re}\left(F^{\prime\prime} \cdot F^{\ast}\right)}{\left\vert F\right \vert^{2}} + 
							  \frac{\left[\textmd{Im} \left(F^{\prime} \cdot F^{\ast}\right)\right]^{2}}{\left\vert F \right\vert^{4}},
\end{equation*}
where all quantities at the right-hand side should be evaluated at a singular point for which $a = \left\vert F\right\vert = 0$.
Boundedness of this expression is highly exceptional, and a generic perturbation of a function for which it is bounded, will lead to unboundedness.

\subsection{SFB wave field} 		\label{SubsectionSFB}

The NLS equation has many interesting special solutions. 
{\spaceskip 0.55 em \relax One family is the so-called `soliton on finite background',} SFB~\cite{34Akhmediev97}. 
This is a remarkable family of solutions since they describe the full nonlinear evolution of the linear instability of surface water waves that is called after Benjamin--Feir~\cite{35Benjamin67}, and as such they are well-suited to study modulational instability in full detail, and to use as a model for generating extreme waves in a hydrodynamic laboratory~\cite{36Andonowati07,37Karjanto06}. 
We describe the main characteristics, using normalized parameters $\beta = \gamma = 1$ and a normalized background

The `background' is a uniform wave train, which corresponds to the solution of NLS given by $A=e^{-i\xi}$.  
Then SFB can be written in the form~\cite{38vanGroesen06}
\begin{equation}
A(\xi,\tau) = e^{-i \xi} \cdot [G(\xi,\tau)e^{i\phi(\xi)} - 1],
\end{equation}
where $G(\xi,\tau)$ and $\phi(\xi)$ are real `displaced' amplitude and phase variables. 
The phase $\phi$ is a monotone function of $\xi$ given by $\phi(\xi) =\arctan\left[-({\sigma}/{\nu}^{2}) \tanh(\sigma \xi) \right]$, where ${\sigma} = {\nu} \sqrt{2 - {\nu}^{2}}$ is the growth rate corresponding to the Benjamin--Feir instability~\cite{35Benjamin67} and $\nu$ is the modulation frequency with $0 <{\nu} < \sqrt{2}$. 
At each fixed position (phase) the dynamics $\tau \mapsto G(\phi, \tau)$ is described by an oscillator equation and given explicitly by $ G(\xi,\tau) = P(\xi)/[Q(\xi) - \sigma \cos(\nu \tau)]$, with $P(\xi) = {\nu} \sqrt{2} \sqrt{2{\nu}^{2} \cosh^{2}(\sigma \xi) - {\sigma}^{2}}$, and $Q(\xi) = {\nu} \sqrt{2} \cosh(\sigma \xi)$.

The fact that $\phi$ is independent of $\tau$ means that at each position the trajectory in the Argand diagram is on a straight line through the point $-1$ under an angle $\phi$. 
Hence, only when $\phi=0$, which means at $\xi=0$, there can be a singular point. 
At that position singular points will occur if $\cos(\nu \tau) = 2(1 - {\nu}^{2})/\sqrt{4 - 2{\nu}^{2}}$. 
Vanishing amplitude occurs for $0 < {\nu} \leq \sqrt{3/2}$ for which there is phase singularity. 
Such phase singularity occurs at $\xi=0$ for two instants in each temporal period. 
At the phase singularities, the local wavenumber and local frequency become unbounded, as shown in Figure~\ref{SFB_LWLF}; this confirms the fact that the Chu--Mei quotient is unbounded at the singular points.
\begin{figure}[htbp]
\begin{center}
\includegraphics[width = 0.35\textwidth]{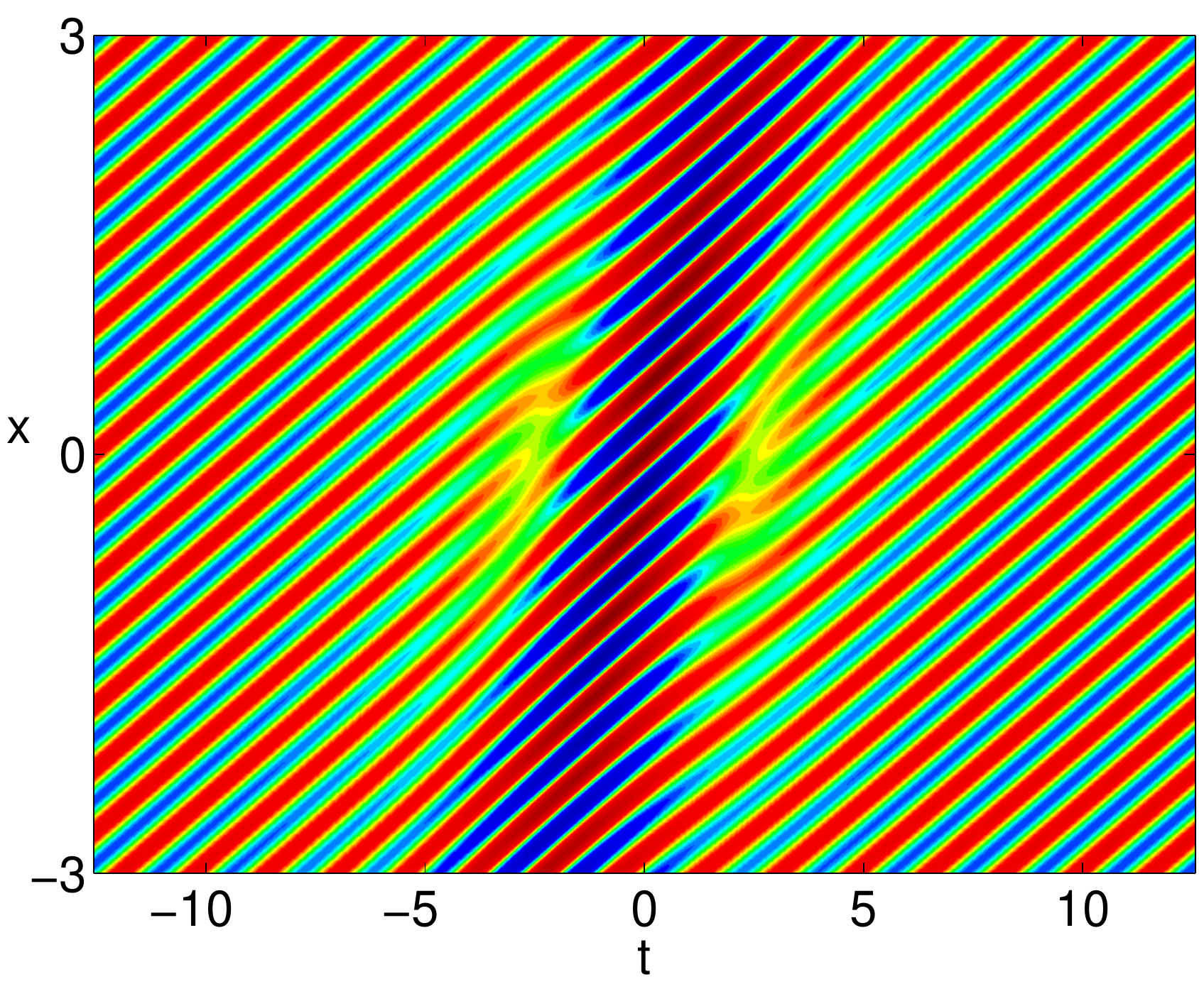}\\	
\includegraphics[width = 0.35\textwidth]{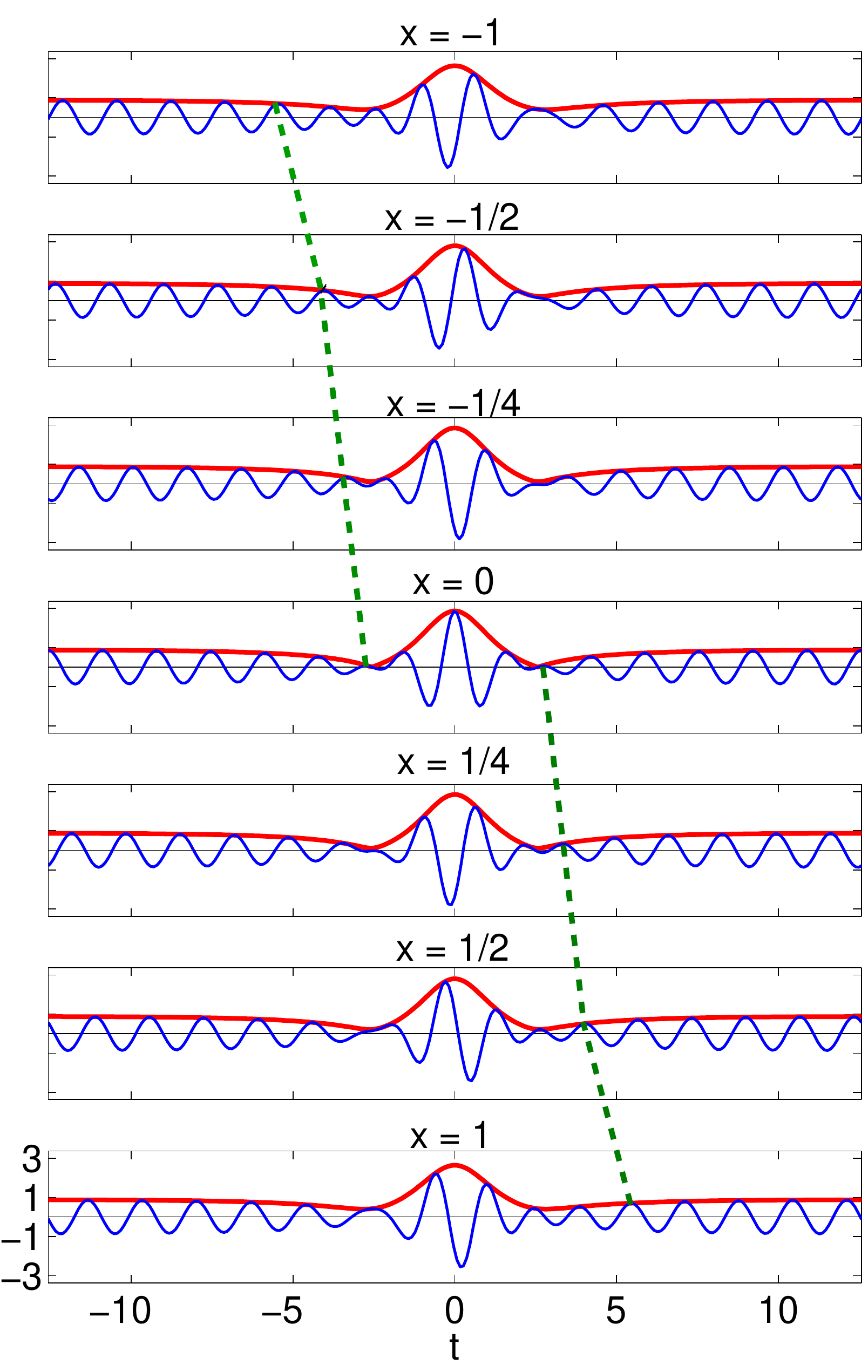} 
\caption{\footnotesize  Density plot of the SFB wave field with wavefront dislocations at the top, and the corresponding wave signals for different positions in a moving frame of reference at the bottom for $\nu = \frac{1}{2}$. 
The number of waves decreases from 8 to 7 for the half period $t \in [-\frac{1}{2}T,0]$ and it increases from 7 to 8 for the next half period $t \in [0,\frac{1}{2}T]$.}  \label{SFB}
\end{center}
\end{figure}

At the singular points there are wavefront dislocations.
Calculation of the strength of the singular points shows that in each period one singular point shows merging, the other one splitting of waves. 
Figure~\ref{SFB} shows a density plot of the SFB wave field around two phase singularities. 
We observe the splitting and merging waves in pairs. 
In plots of the time signal at different positions, we see the splitting and merging in more detail. 
In this example, for half modulation period $t \in [-\frac{1}{2}T,0]$, the number of waves decreases from 8 to 7, indicating that waves are merging when passing the singularity. 
At another half modulation period $t \in [0,\frac{1}{2}T]$, it increases from 7 to 8, which indicates that waves are splitting when passing the singularity. 
However, the number of waves in one modulation period for $x \rightarrow \pm \infty$ remains the same before and after undergoing the singularity, namely $\omega_{0}/\nu$. 
For a more detailed discussion of the SFB and related special solutions of NLS on finite background see~\cite{36Andonowati07,37Karjanto06,38vanGroesen06}.

An observation and investigation of wavefront dislocation in modulated surface water waves has been done by Tanaka~\cite{7Tanaka95}. 
His investigation is based on the modulated gravity waves corresponding to Benjamin--Feir instability and is done numerically. 
By taking an analogy to our signalling problem, the corresponding envelope function experiences vanishing amplitude at two different positions. 
He observed that between these two vanishing amplitudes, the wave crests `disappear', as is confirmed by the decrease in the number of waves.

\section{Conclusions}

We discussed the phenomena of phase singularity and wavefront dislocation that can happen at singular points of a wave field where the amplitude vanishes. 
We used simple examples of trichromatic waves to see the relationship between these concepts. 
We also linked the unboundedness of the Chu--Mei quotient to the unboundedness of the local wavenumber and frequency at singular points. 
It is important to stress again that the phenomena are essentially linear since nonlinear terms in the equation are of higher order at a singular point. 
We showed that for an interesting class of solutions of the NLS equation, the solitons on finite background, that wavefront dislocations occur there too.

\section*{Acknowledgements} 
This work is executed at the University of Twente, The Netherlands as part of the project \textit{``Prediction and generation of deterministic extreme waves in hydrodynamic laboratories"} (TWI.5374) of the Netherlands Organization of Scientific Research NWO, subdivision Applied Sciences STW.

{\small \balance 

}

\begin{thebibliography}{99}

\bibitem{1Balistreri00} M.L.M. Balistreri, J.P. Korterik, L. Kuipers, N.F. van Hulst, Phys. Rev. Lett. 85 (2000) 294.

\bibitem{2Nye99} J.F. Nye, Natural Focussing and Fine Structure of Light: Caustics and Wave Dislocations, IOP, Bristol, 1999.

\bibitem{3Berry00} M.V. Berry, M.R. Dennis, Proc. R. Soc. London A 456 (2000) 2059.

\bibitem{4Berry01} M.V. Berry, M.R. Dennis, Proc. R. Soc. London A 457 (2001) 2251.

\bibitem{5Dennis04} M.R. Dennis, J. Opt. A: Pure Appl. Opt. 6 (2004) S155.

\bibitem{6Soskin01} M.S. Soskin, M.V. Vasnetsov, Prog. Opt. 42 (2001) 219.

\bibitem{7Tanaka95} M. Tanaka, in: Proceedings IUTAM/ISIMM Symp. Struct. Dynamics Nonli. Waves Fluids, 1995, p. 392--398.

\bibitem{8Orowan34} E. Orowan, Z. Phys. 89 (1934) 605; \\ E. Orowan, Z. Phys. 89 (1934) 614; \\ E. Orowan, Z. Phys. 89 (1934) 634.

\bibitem{9Polanyi34} M. Polanyi, Z. Phys. 89 (1934) 660.

\bibitem{10Taylor34} G.I. Taylor, Proc. R. Soc. London 145A (1934) 362.

\bibitem{11Hull01} D. Hull, D.J. Bacon, Introduction to Dislocations, fourth ed., Butterworth Heinemann, Oxford, 2001.

\bibitem{12Dieter88} G.E. Dieter, Mechanical Metallurgy, SI Metric ed., McGraw--Hill, London, 1988.

\bibitem{13Honeycombe84} R.W.K. Honeycombe, The Plastic Deformation of Metals, second ed., Edward Arnold, London, 1984.

\bibitem{14Read53} W.T. Read Jr., Dislocations in Crystals, McGraw--Hill, New York, 1953.

\bibitem{15Dennis01} M.R. Dennis, Ph.D. Thesis, University of Bristol, 2001.

\bibitem{16Krasa81} J. Kr\'{a}sa, J. Phys. D: Appl. Phys. 14 (1981) 1241.

\bibitem{17Berry80} M.V. Berry, et al., Eur. J. Phys. 1 (1980) 154.

\bibitem{18Wright79} F.J. Wright, in: W. G\"{u}ttinger, H. Eikemeier (Eds.), Structural Stability in Physics, Springer, Berlin, 1979, pp.~141--156.

\bibitem{19Basistiy95} I.V. Basistiy, M.S. Soskin, M.V. Vasnetsov, Opt. Commun. 119 (1995) 604.

\bibitem{20Nabarro04} F.R.N. Nabarro (Ed.), Dislocations in Solids 1--12 (1979-2004).

\bibitem{21Nye97} J.F. Nye, Proc. R. Soc. London A 355 (1997) 2065.

\bibitem{22Nye81} J.F. Nye, Proc. R. Soc. London A 378 (1981) 219.

\bibitem{23Nye&Berry74} J.F. Nye, M.V. Berry, Proc. R. Soc. London A 336 (1974) 165.

\bibitem{24Trulsen98} K. Trulsen, J. Geophys. Res. 103 (1998) C2:3143.

\bibitem{25Berry81} M.V. Berry, Les Houches 1980, Session XXXV, Physics of Defects, North-Holland, Amsterdam, 1981, pp.~453--459.

\bibitem{26Berry98} M.V. Berry, in: Proceedings of International Conference on Singular Optics, SPIE, 1998, pp.~1--10.

\bibitem{27Infeld90} E. Infeld, G. Rowlands, Nonlinear Waves, Solitons and Chaos, Cambridge Univ. Press, Cambridge, 1990, pp.~117--119.

\bibitem{28Fornberg78} V. Fornberg, G.B. Whitham, Phil. Trans. R. Soc. London A 289 (1978) 373.

\bibitem{29Chu70} V.H. Chu, C.C. Mei, J. Fluid Mech. 41 (1970) 873.

\bibitem{30Chu71} V.H. Chu, C.C. Mei, J. Fluid Mech. 47 (1971) 337.

\bibitem{31Whitham67} G.B. Whitham, J. Fluid Mech. 27 (1967) 399.

\bibitem{32Karpman67} V.I. Karpman, JETP Lett. 6 (1967) 277.

\bibitem{33Karpman69} V.I. Karpman, E.M. Krushkal', Sov. Phys. JETP 28 (1969) 277.

\bibitem{34Akhmediev97} N.N. Akhmediev, A. Ankiewicz, Solitons--Nonlinear Pulses and Beams, Chapman and Hall, London, 1997 pp.~50--57.

\bibitem{35Benjamin67} T.B. Benjamin, J.E. Feir, J. Fluid Mech. 27 (1967) 417.

\bibitem{36Andonowati07} Andonowati, N. Karjanto, E. van Groesen, Appl. Math. Mod. 31 (2007) 1425.

\bibitem{37Karjanto06} N. Karjanto, Ph.D. Thesis, University of Twente, 2006.

\bibitem{38vanGroesen06} E. van Groesen, Andonowati, N. Karjanto, Phys. Lett. A 354 (2006) 312.
\end{thebibliography}
\end{document}